\documentclass[12pt]{article}
\usepackage{float}
\usepackage{amsmath}
\usepackage{amsfonts}
\usepackage{amssymb}
\usepackage{graphicx}
\usepackage{cite}
\usepackage{color}
\usepackage{dcolumn}
\usepackage{wrapfig}
\usepackage{longtable}
\usepackage{times}
\usepackage{caption}
\usepackage{subcaption}
\usepackage{hyperref}
\usepackage{bm}
\usepackage[margin=0.7in]{geometry}
\usepackage{newfloat}
\DeclareFloatingEnvironment[name={SM Figure}]{suppfigure}
\DeclareFloatingEnvironment[name={SM Table}]{supptable}
\begin{document}
\begin{flushleft}
{\LARGE
\textbf{Taming Chimeras in Networks {through} Multiplexing Delays}
}
\\
\vspace{0.3cm}
\bf Saptarshi Ghosh$^1$,Leonhard Sch\"ulen$^3$,Ajay Deep Kachhvah$^1$, Anna Zakharova$^3$ and Sarika Jalan$^{1,2,\ast}$
\\
\vspace{0.2cm}
\it ${^1}$ Complex Systems Lab, Discipline of Physics, Indian Institute of Technology Indore, Khandwa Road, Simrol, Indore 453552, India
\\
\it ${^2}$ Discipline of Biosciences and Biomedical Engineering, Indian Institute of Technology Indore,  Khandwa Road, Simrol, Indore 453552, India
\\
\it ${^3}$ Institut f{\"u}r Theoretische Physik, Technische Universit\"at Berlin, Hardenbergstra\ss{}e 36, 10623 Berlin, Germany
\\
${^{\ast}}$ Corresponding Author, email: sarika@iiti.ac.in
\end{flushleft}

\begin{abstract}
	Chimera referring to a coexistence of coherent and incoherent states, is traditionally very difficult to control due to its peculiar nature. Here, we provide a recipe to construct chimera states in the multiplex networks with the aid of multiplexing-delays. The chimera state in multiplex networks is produced by introducing heterogeneous delays in a fraction of inter-layer links, referred as multiplexing-delay, in a sequence. Additionally, the emergence of the incoherence in the chimera state can be regulated by making appropriate choice of both inter- and intra-layer coupling strengths, whereas the extent and the position of the incoherence regime can be regulated by appropriate placing and {strength} of the multiplexing delays. The proposed technique to construct such {engineered} chimera equips us with multiplex network's structural parameters as tools in gaining both qualitative- and quantitative-control over the incoherent section of the chimera states and, in turn, the chimera. Our investigation can be of worth in controlling dynamics of multi-level delayed systems and attain desired chimeric patterns.

\end{abstract}

\section{Introduction} Chimera, a hybrid dynamical state representing coexistence of coherence and incoherence in a system, is relatively a newly-found partial synchronized state~\cite{chim.def,chim.multiplex} and has extensively been investigated both theoretically~\cite{chim.discrete.cont,chim.discrete.cont_2,chim.discrete.cont_3,chim.discrete.cont_4} and experimentally~\cite{chim.expp} for a diverse range of natural and artificial complex systems represented by complex networks~\cite{chim.review}. Chimera-like patterns have been detected in different parts of the human brain~\cite{chim.neuro.rev} which interact among themselves to perform different cognitive tasks~\cite{chim.brain_network}. Intracranial Electroencephalography readings also have detected chimera-like patterns at the onset of epileptic seizures~\cite{EEG_chim}. Moreover, the emergence of the chimera is also found to be driven by the presence of inhibition which is ingrained in the brain network~\cite{chim.inhi}. Subsequently, it is also important to gain control over chimera state whose characteristics can be guided~\cite{chim_control}. Recently, it has been shown that one can produce tailor-made chimera by inducting heterogeneous delays in the edges of a network~\cite{eng.chim}.
Furthermore, Multilayer networks representing various complex systems with multilevel-interactions have fascinated many researchers \cite{lm,mm} and lately have become an important test-bed for the study of various processes such as chimera~\cite{chim.multiplex}, solitary state~\cite{MIK18} ,synchronization~\cite{multi_appl}, percolation~\cite{preloc.multiplex}, coherence resonance~\cite{SEM18}, traffic-transport~\cite{transport.multiplex} etc. A multiplex network basically is a layered representation of a complex system having multilevel interactions, in which interacting units are the nodes having multiple types of interactions among them, with each type of interaction creating one layer~\cite{mul_def}. 
For instance, two cities connected by multiple modes of transportation (bus, train, and flight), and the same set of neuronal blocks interacting in different ways to perform different tasks~\cite{mul_rev}. 
Additionally, delayed interactions are known to exert substantial influence on several emergent phenomena such as synchronization~\cite{delay_sync}, oscillation death~\cite{delay_osc_death}, chimera states~\cite{delay_chim,SAW17,SAW19a,GJU17}, in the coupled dynamical systems. Due to the finite speed of information transmission, delay naturally arises between the transmission ends (nodes) connected through channels (links)~\cite{delay_book}. For instance, in a neural network represented by a multiplex framework where intralayer and interlayer links denote electrical and chemical synapses, respectively, the longer reaction time of the chemical synapses might induce a delay in the interlayer links\cite{delay_neu}.
In a similar fashion, in a multi-modal transport network where different modes of transportation are denoted by different layers~\cite{delay_trans}, a delay may arise because of passengers switching the mode of transportation.
Motivated by the feasibility of the delayed multiplex networks in mimicking natural systems, {we investigate engineering of chimeric patterns in such a framework. Previous studies had demonstrated an interplay of system-wide inter- and intra-layer delay can help in emergence of chimera state in multiplex networks~\cite{SAW17,delay_chim}}.  
{However, in} this article, we present a strategy to engineer chimera in a multiplex network by installing {heterogeneous} delays in {certain} inter-layer links. {It has been reported that a chimera state can be instated by means of heterogeneous delays in intra-layer links of a monoplex network such that the position and extent of incoherent region(s) is completely controlled by placement and values of the delays,respectively~\cite{eng.chim}. Here, we display} that the chimera state can be {induced through such schemes} in originally completely coherent identical multiplexed layers by installing heterogeneous delays in a {fraction} of interlayer links. {We demonstrate that the incoherent section of the chimera, hence, in turn, the chimera can be regulated both qualitatively and quantitatively only with combined aid of multiplex networks' structural parameters and a fraction of multiplexing delays, respectively. We also show that such} manufactured chimera is independent of initial conditions, which is otherwise conventionally mandatory for the existence of chimera in coupled maps.
\begin{figure}
      \centerline{\includegraphics[width=0.8\columnwidth]{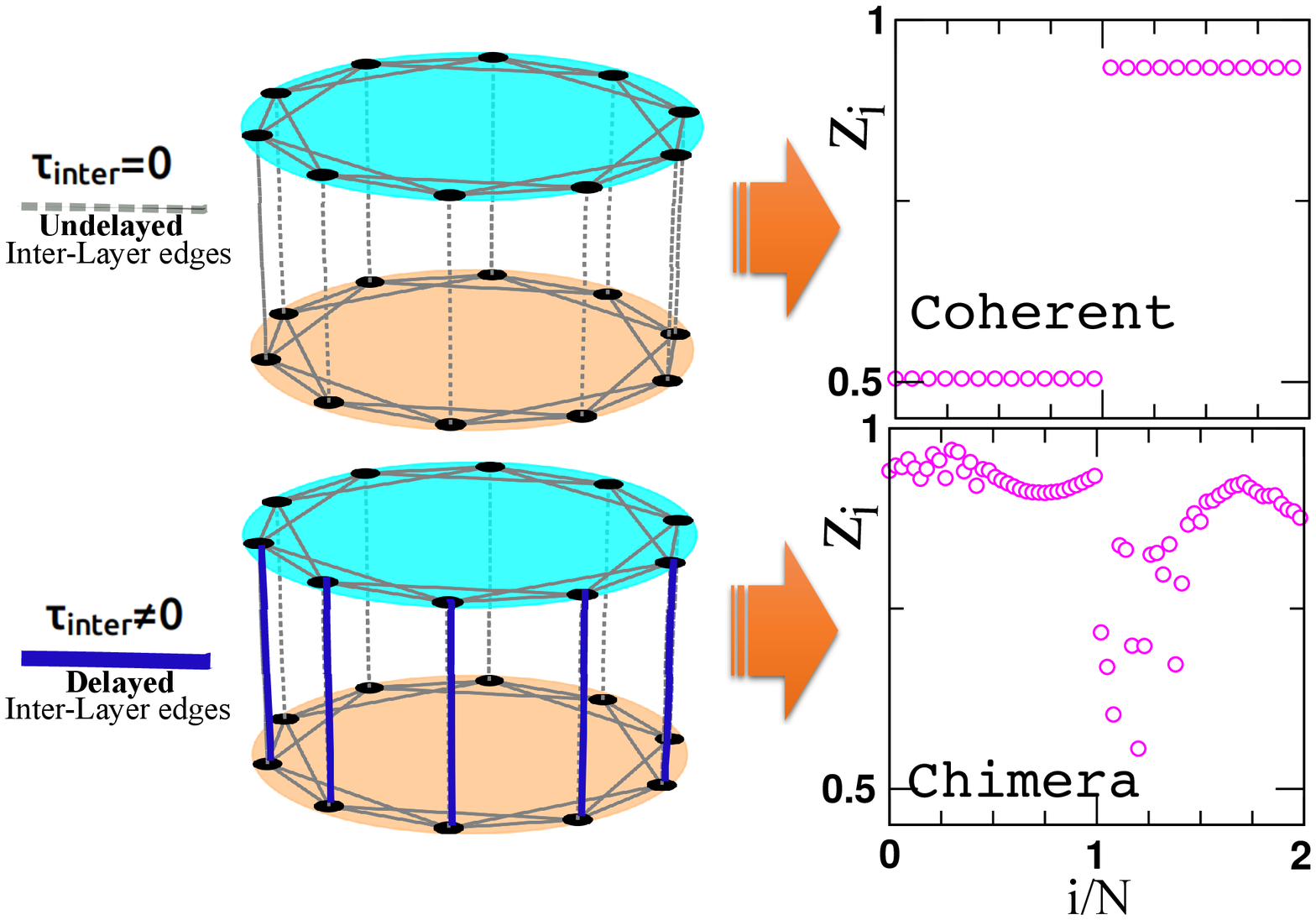}}
      \caption{(Color online) Schematic diagram of multiplex networks (left panel) comprising two identical regular networks  is shown with un-delayed (top panel) and delayed (bottom panel) interlayer links. Right panel illustrates that the undelayed interlayer links give rise to completely coherent states in both the layers, whereas the delayed interlayer links produce chimera state, a mixture of coherence and incoherence behavior in both the layers where incoherence is facilitated by the delayed interlayer links.}
      \label{fig1}
	  \end{figure}

\section{Model} In the current work, we focus on demonstrating chimera in multiplex networks, arising due to distinct time-delays present in a fraction of the interlayer links. To achieve this, we consider an undirected multiplex network constructed from two identical regular networks ($S^1$: ring), each having $N$ nodes. Two layers of the multiplex network are encoded by a set of adjacency matrices $\{A_1,A_2\}$, hence multiplex network $\mathcal{A}$ can be expressed as 
\begin{equation}
	\mathcal{A}=
	\begin{pmatrix} A_1 & D_xI \\ D_xI & E_yA_2 \end{pmatrix},
	\label{mul_mat}
\end{equation}
where $I$ is an identity matrix representing links between one-to-one mirror nodes in two layers. Matrix element ${A}_{ij}=1$, if a link exists between nodes $i$ and $j$, and ${A}_{ij}=0$, otherwise. The parameter $D_x$ represents multiplexing strength by which a node and its counterpart in another layer impact each other's dynamics. The parameter $E_y$ depicts a scaling parameter for intralayer coupling strength of a layer \cite{mul_net_weighted}. 
We define coupling matrix $C=\varepsilon \mathcal{A}$, where its element $C_{ij}$ denotes the effective coupling strength between $i^{th}$ and $j^{th}$ nodes with $\varepsilon\in[0,1]$ representing the overall coupling strength. A schematic diagram of the multiplex network 
is shown in Fig.~\ref{fig1}.

The local dynamics of each node in the multiplex network is represented by a logistic map $f(z)=\mu z (1-z)$ where bifurcation parameter $\mu\in[0,4]$ and state ${z}(t)\in[0,1]$ ~\cite{logsMap}. We have considered the map in its chaotic regime ($\mu=4$). The dynamical evolution of the nodes in the multiplex network is governed by~\cite{delay_sync}
\begin{equation}
z_i(t+1)=f(z_i(t))+\frac{1}{(k_i)} \sum_{j=1}^{2N} C_{ij}[ f(z_j(t-\tau_{ij}))-f(z_i(t)) ]
\label{eq.evol}
\end{equation}
where $i=1,\ldots,2N$, normalizing factor $ k_{i}$ = $\sum_{j=1}^{2N}C_{ij}$ and parameter $\varepsilon\in[0,1]$ is the homogeneous coupling constant. We furthermore introduce delay in the dynamics by delay matrix $\tau$ whose symmetric element $\tau_{ij}=\tau_{ji}$ represents delay between $i^{th}$ and $j^{th}$ node. We choose a fraction $N_{\tau}$ of the interlayer links in a sequence and each chosen link is then assigned a delay value selected uniformly randomly in the range $0\le\tau_{ij}\le\tau_{max}$~\cite{eng.chim}. The symmetric upper right and lower left blocks $I$ of the adjacency matrix $\mathcal{A}$ possess the interlayer delayed links.

The chimera state corresponds to a {hybrid state with presence} of both coherent and incoherent dynamics. Next, we define the criteria for the existence of {(in)}coherence in the coupled maps as follows~\cite{chim.discrete.cont}
\begin{equation}
\lim\limits_{N \rightarrow \infty} \lim\limits_{t \rightarrow \infty}\sup\limits_{i,j  \in U_{\xi}^N (x)} \mid{z_i(t)-z_j(t)} \mid \rightarrow 0 \, \, \text{for} \, \, \xi\; \rightarrow 0
\label{eq.cohr}
\end{equation}
where $U_{\xi}^{N} (x) = \{ j : 0 \le j \le N, \mid{\frac{j}{N} - x} \mid <  \xi \}$ represents the {neighborhood} of a node {in regular(ring) network ($x \in S^1$). Thus, the state $z(x, t)$ assumes a profile such that all the nodes possess low spatial distance with their neighbors, approaching a smooth spatial curve of a coherent state in the asymptotic limit of $N\to\infty$.} Any break in the profile {,i.e., high spatial distance in neighboring nodes} is depicted as the incoherence~\cite{chim_CSF}. Therefore, the snapshots or the spatial curves  (schematically presented in Fig.~\ref{fig1}; right bottom corner) depict a chimera state if a smooth region {(coherent part with closely placed neighboring nodes)} coexist with a region characterized by scattered points {(incoherent part with distantly placed neighbors). Furthermore, a complete coherence can be attributed to the state $z$ when all nodes assume same constant value with zero spatial distance in neighbors, thus producing a straight spatial curve.} 
\begin{figure}[t]
	\centerline{\includegraphics[width=0.8\columnwidth]{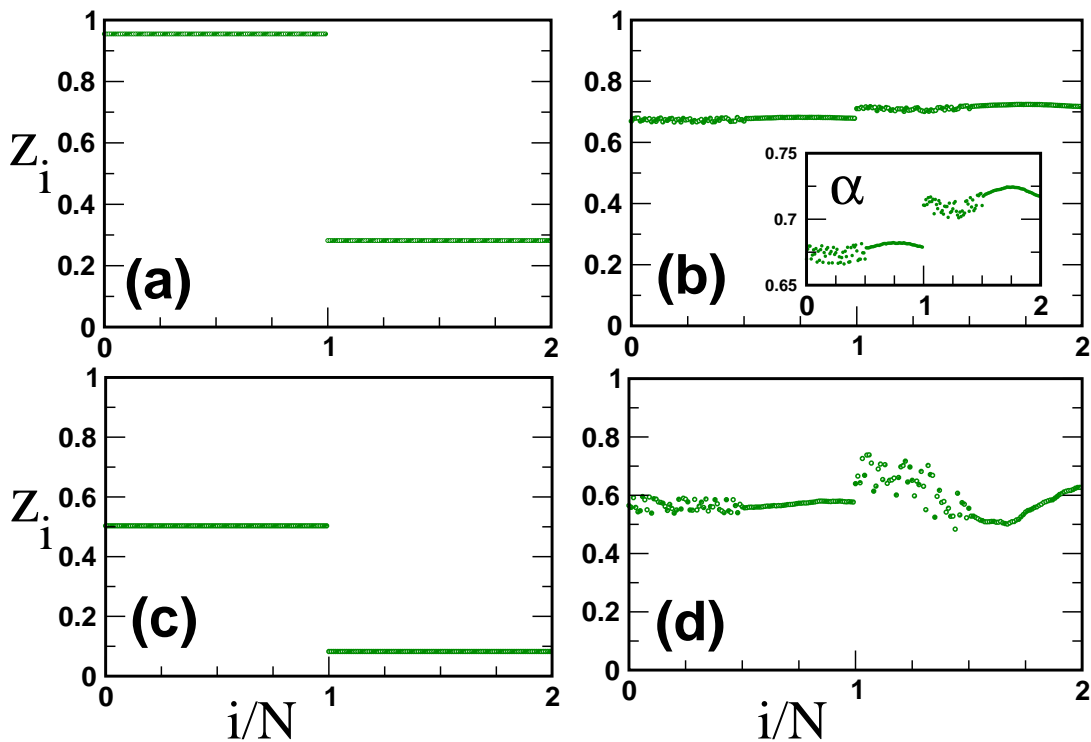}}
	\caption{(Color online) Snapshot profiles of the two layers of the considered multiplex network for (a) the undelayed interlayer links, (b) the delayed interlayer links with parameters $D_x=1, E_y=1$ and similarly, (c) the undelayed interlayer links, and (d) the delayed interlayer links with parameters $D_x=1, E_y=0.6$. The inset curve (b($\alpha))$ displays the slight disturbance in (b) in magnified Y axis. The parameters are $\varepsilon=0.9$, $\tau_{max}=20$, $r=0.32$, $N=100$ in each layer.
	}
	\label{fig2}
	\end{figure}

\section{Results} We investigate emergence of chimera states due to the presence of multiplexing-delays in multiplex networks. Furthermore, we demonstrate how the emergent chimera can be regulated by means of structural parameters of the multiplex network. \newline
\subsection{Chimeric patterns induced by multiplexing delays.-}
To carry out various types of numerical analysis, we, throughout the paper, consider a multiplex of two ring-regular networks, each having $N=100$ nodes with coupling radius $r=0.32$, same in both layers i.e. all the nodes are connected with its $64$ neighbors (Node degree $\langle k \rangle = 2Nr = 64$)\cite{chim.discrete.cont}. For all the simulations, we introduce heterogeneous delays $\tau\in(0,\tau_{max}=20]$ drawn from a uniform random distribution, at half of the inter-layer links $N_\tau=N/2$ chosen sequentially, otherwise mentioned elsewhere. 
{An identical set of initial states for the two layers give rise to identical states for both the layers~\cite{chim.multiplex}. Hence to demonstrate the robustness of our technique we have opted two distinct sets of initial states for the maps, which are selected randomly $z\in[0,1]$ for the two multiplexed layers.}
The system of networked maps is updated for a sufficiently large time $5\times10^4$ and the snapshot of final states of all the nodes for both the layers is recorded. The coupling parameter is kept fixed at $\varepsilon=0.9$, so that the system of networked maps initially remains in the coherent state as shown in the Fig.\ref{fig2}(a) when there is no delay present in the system ($\tau_{i,j}=0 ;\forall i , \forall j$).
Fig.~\ref{fig2}(b) depicts the dynamical profile of the multiplex network in which induction of delays leads to a slight disturbance (Fig.~\ref{fig2}(b); inset $\alpha$ ) in the pattern of coherent state of the nodes. However, no chimera pattern is observed in this case. The reason is that both the multiplexed layers are dense networks and the mirror nodes which are connected by delayed interlayer links are also connected to a large number of neighbors by undelayed intralayer links. Therefore, the delayed mirror nodes fail to get completely separated from the rest of the coherent nodes in both the layers. This situation can be made to favor the emergence of chimera by varying structural parameters in such a way that the perturbative (incoherent) effects arising from delayed interlayer links become more dominant. Fig.~\ref{fig2}(d) depicts that the contribution of the delayed interlayer links can be enhanced by setting $D_x$ and $E_y$ appropriately. However, Fig.~\ref{fig2}(c) displays a coherent dynamical profile with the same of $D_x$ and $E_y$ values (as Fig.\ref{fig2}(c)), but without the multiplexing delays. Thus, a combination of a high value of $D_x$ and a very low value of $E_y$ is perfect, along with the multiplexing delays, to obtain chimera state. In Fig.~\ref{fig2}(d), chimera state emerges in the second layer, while the nodes in the first layer experience only {faint disturbance}. Thereby, this recipe helps us to attain complete regulatory control over emergence of chimeric patterns in both the layers by suitably choosing $D_x$ \& $E_y$ values. Additionally, we can entirely suppress the chimera  inducted by appropriate placement of the multiplexing delays, in one layer while having the desired chimera pattern in another layer by tuning the $D_x$ and $E_y$ values. \newline
\subsection{Role of the $D_x$ \& $E_y$ parameters.-}
\begin{figure}[t]
	\centerline{\includegraphics[width=0.8\columnwidth]{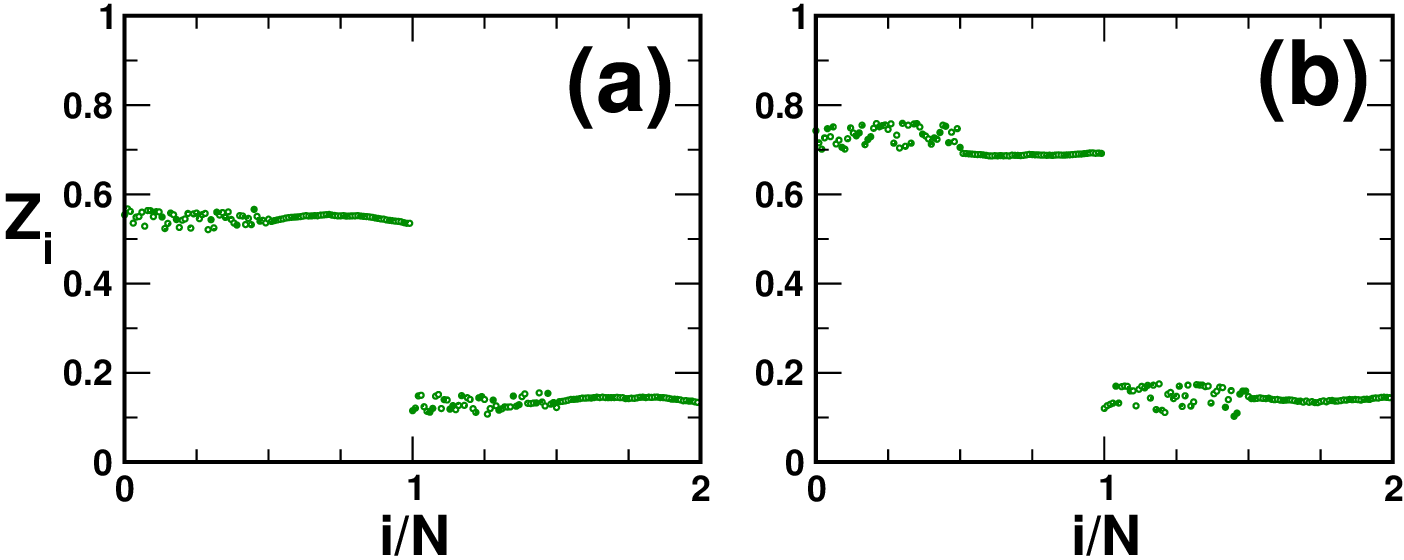}}
	\caption{(Color Online) Snapshots of both the layers of the multiplex network where half of the inter-layers edges are heterogeneously delayed for the multiplexing parameters (a) $D_x=3,E_y=1$, (b) $D_x=5,E_y=1$. Others are same as described in Fig.\ref{fig2}.}
	\label{fig3}
	\end{figure}

Next, we take a closer look on the role of multiplex network's structural parameters namely $E_y$ and $D_x$ to understand the collective dynamical behavior of the layers. Fig.~\ref{fig3}, represents chimera states for different choices of $D_x $ with usual scaling value of intralayer coupling strength $E_y=1$. Fig~\ref{fig2}(b) shows that the usual choice of $D_x=1,E_y=1$ yields a slight wobbling in the nodes connected with the delayed inter-layer links. A high value of $D_x$ changes the situation drastically as can be seen in Fig.~\ref{fig3}. High value of $D_x$ causes an enhancement in the connection strength between each pair of the mirror nodes permitting the interlayer links dominating over the intralayer links in influence. Thus, the mirror nodes {disperse} more freely in both the layers because of the multiplexing delays. Fig.~\ref{fig3}(a) shows an increment in the {dispersal of nodes connected to delayed inter-layer links} (thus, producing engineered chimera states) in both the layers due to higher values of $D_x(=3)$. Increased $D_x$ induce {even larger spatial separation} for the mirror nodes in both the layers even with high value of $E_y(=1)$. An even more noticeable chimera is observed in Fig.~\ref{fig3}(b) as the value of $D_x(=5)$ is higher in this case. 
	\begin{figure}[t]
		\centerline{\includegraphics[width=0.8\columnwidth]{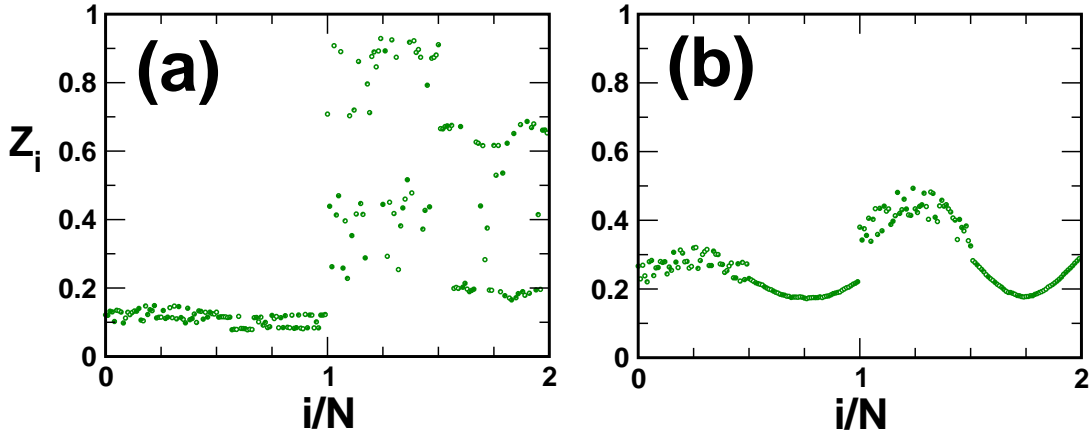}}
		\caption{(Color Online) Snapshots of both the layers of the multiplex network where half of the inter-layers edges are heterogeneously delayed for the multiplexing parameters (a) $D_x=5,E_y=0.2$, (b) $D_x=5,E_y=0.8$. Others are same as described in Fig.\ref{fig2}.}
		\label{fig4}
	\end{figure}
\begin{figure*}
\begin{minipage}[c]{0.67\textwidth}
\includegraphics[width=\textwidth]{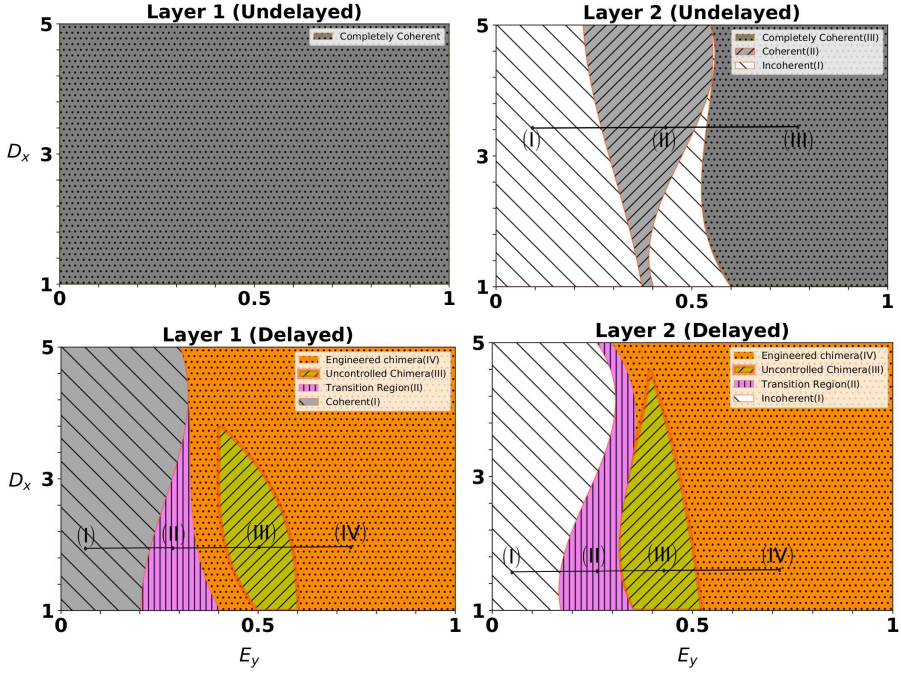}
\end{minipage}\hfill
\begin{minipage}[c]{0.3\textwidth}
\caption{(Color Online) A schematic phase diagram of $D_x$ - $E_y$ space for (a) layer 1 and (b) layer 2 with undelayed inter-layer edges and (c) layer 1 and (d) layer 2 with delayed inter-layer edges. Half of the inter-layers edges are heterogeneously delayed (in c \& d) and the delay values are chosen from a uniform random distribution with $\tau_{max}=20$. The parameters are same as Fig.~\ref{fig2}. The {boundaries of various regions are drawn} from visual inspection and measures (See SM~\cite{SM}).}
\label{fig5}
\end{minipage}
\end{figure*}
Next, we present the impact of $E_y$ along with a fixed high value of $D_x(=5)$ in Fig.~\ref{fig4}.  A small value of $E_y$ dilutes the intralayer contributions among neighboring nodes in the corresponding layer allowing the nodes to {strew} more freely under the influence of the multiplexing delays. In Fig.~\ref{fig4}(a), the mirror nodes experiencing the multiplexing delays, {show low spatial separation in neighboring nodes} in the first layer, whereas their counterparts in the second layer experience a complete incoherent state due to very weak $E_y$. The {dispersal} of the nodes in the second layer can be tamed by increasing $E_y$ (see Fig. ~\ref{fig2}(d) and Fig.~\ref{fig4}(b)). Increment in $E_y$ diminishes the relative difference between the intralayer coupling strengths of both the layers. Note that Fig. ~\ref{fig3} (b) presents a well pronounced chimeric spatial profile, whereas Fig.~\ref{fig2} (b) shows a slight wobble in the mirror nodes experiencing multiplexing delays, although values of $E_y$ are same in both cases. Thus observed chimera in the second layer can be made even more pronounced by fine-tuning $E_y$ while keeping $D_x$ high (see Fig.~\ref{fig4} (b)). From these observations it is apparent that $D_x$ helps in introducing incoherence in the mirror nodes in both the layers by means of multiplexing heterogeneous delays, whereas $E_y$ essentially brings in incoherence in a layer by diminishing coupling intensity of intralayer links. 
 The interplay between these two parameters can give rise to the emergence of pronounced chimera in only one layer while the mirror nodes in another layer can cause only mild disturbance. Hence, the intensity of chimera states in a multiplex network can be regulated by inducting multiplexing heterogeneous delays with appropriate choices of the network's structural attributes. 
{So far, we have demonstrated the existence of chimera for a few combinations of the parameters $D_X$ and $E_y$, though the control scheme is applicable for a wide range of values of the parameters. Fig.~\ref{fig5} presents phase diagrams in $D_{x} - E_{y}$ space exploring different emerging states including chimera in the multiplex network both in the absence and the presence of the inter-layer delays. Here the phase $D_{x} - E_{y}$ diagrams correspond to a high coupling strength ($\varepsilon=0.9$) so as to have the synchronous clusters in the multiplexed layers, which could be perturbed to explore the existence of chimera by incorporating multiplexing delays. Note that the schematics and boundaries of phase diagrams in Fig.~\ref{fig5} are based on the variance~\cite{var_sync} (upper panels; Fig.~\ref{fig5}(a,b)) and correlation measure~\cite{kemeth} ($g_0(t)$; lower panels; Fig.~\ref{fig5}(c,d)) defined and discussed in the supplementary material~\cite{SM}.
Panels~\ref{fig5} (a) and \ref{fig5} (b) show the coherence profile for Layer 1 and Layer 2, respectively, in the absence of multiplexing delays. Layer 1 displays completely coherent states spanning the entire $D_{x}-E_{y}$ space due to high coupling strength whereas Layer 2, due to the effective coupling strength $E_y*\varepsilon$, shows coherent states (regime II) in the mid-range of $D_{x}-E_{y}$ space and completely coherent states (regime III) for $E_y>0.5$.
Panels~\ref{fig5} (c) and \ref{fig5} (d) exhibit chimera profiles for Layer 1 and Layer 2, respectively, when the inter-layer delays are present. Layer 1 shows coherent region (regime I) while Layer 2 shows incoherent region (regime I) for low values of $E_y$ and engineered chimera states (regime IV) for mid- and high-range values of $E_y$. Regime III for both the layers depicts un-controllable chimera in a sense that the shape or area of the incoherence can not be tailored to one’s preference under this parameter regime. Transition region (regime II) in both the layers yields unidentified states qualified to be neither chimera nor incoherent states. The difference in the effective coupling strength for Layer 1 ($\varepsilon$) and Layer 2 ($E_y*\varepsilon$) accounts for the different $D_{x}-E_{y}$ ranges of regime II and regime III for the two layers. Note that a distinct uniformly colored pattern is used to represent each region in $D_{x}-E_{y}$ diagrams, hence the dotted pattern does not show the qualitative or quantitative variation in the engineered chimera profiles (IV) with change in the value of $D_x$ or $E_y$ (as illustrated in Fig.~\ref{fig3} and Fig.~\ref{fig4}). Hence, the phase diagrams in $D_{x} - E_{y}$ space highlight the importance of network's structural parameters in guiding or regulating the emergent chimera in both the layers of the multiplex network.} \newline
\subsection{Investigating the temporal behavior.-}
\begin{figure*}
	\begin{center}
		\includegraphics[width=7in,height=1.4in]{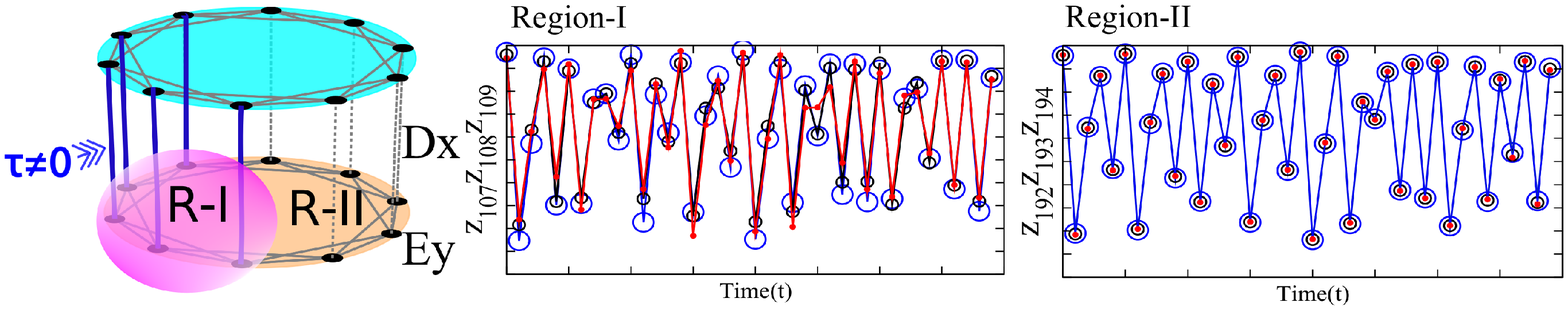}
	\caption{Time series of nodes connected with delayed and undelayed inter-layer edges along with a diagram of a multiplex network consisting of two identical regular network layers. Half of the interlayer edges are heterogeneously delayed (represented by Bold blue lines). Appropriate networking parameters ($D_x = 5$, $E_y = 0.2$) are used to induce chimera in the second layer. Region I (Shaded pink circle) consists of nodes connected with delayed edges and shown to have an incoherent time evolution in the middle panel whereas the nodes of region II (connected to undelayed edges) are shown to have coherent evolution in the rightmost	panel. Together, they display a chimera pattern as shown in Fig.~\ref{fig2}(c). Other parameters are same as in Fig.\ref{fig2}.}
	\label{fig6}
	\end{center}
	\end{figure*}

 A chimera state typically requires a special initial condition for its existence and generally arises in mid-coupling range. In our work, {the perturbation induced in the form of heterogeneous delays at coveted position and length of the sequence of inter-layer links gives rise to the chimera state in the coherent regime.} The occurrence of such engineered chimera in individual layers of the multiplex network is not surprising. The inducted multiplexing delays disturb the respective nodes, in turn, causing dynamical symmetry breaking of the perturbed nodes from the rest of the nodes in the coherent bulk. We also look at the time-evolution of the perturbed (incoherent) and unperturbed (coherent) nodes to get a deeper insight of the chimera state. Fig.~\ref{fig6} shows the time series of six nodes selected from a chimera state, half of the nodes possessing delayed inter-layer links (node index $z_{107}$, $z_{108}$, $z_{109}$), while the rest half possessing undelayed interlayer links (node index $z_{192}$, $z_{193}$, $z_{194}$). The time series of the delayed nodes ($z_{107}$, $z_{108}$, $z_{109}$) shows a desynchronized time evolution as the nodes evolve experiencing different delay values. Nevertheless, the undelayed nodes ($z_{192}$, $z_{193}$, $z_{194}$) maintain their synchronized temporal evolution as they experience no perturbation. 
 {It is important to note that if the interpolated multiplexing delays are homogeneous or identical, the impact of the perturbation will be similar to all disturbed nodes. This will produce synchronous cluster(s), possessing same displacement from the main synchronous cluster, whose displacement in the spatial profile would depend upon the strength of homogeneous delays. However, this kind of spatial profile with detached synchronous clusters can arguably treated either as a cluster synchronized state or as a point-wise chimera states. Therefore, heterogeneous multiplexing delays are better suited for the demonstration of engineered chimera, presented in this study.} \newline
 \subsection{Designing the incoherent region by multiplexing delays.-}
\begin{figure}[!hbp]
	\centerline{\includegraphics[width=0.8\columnwidth]{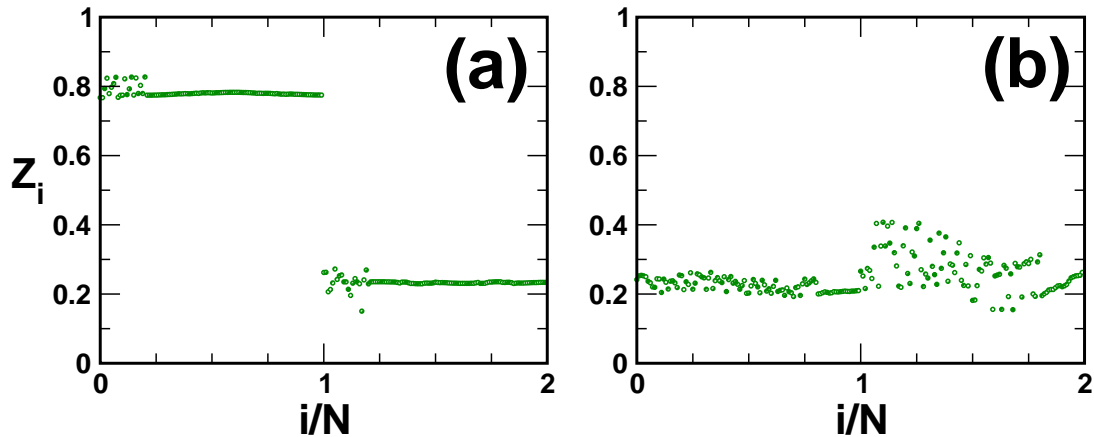}}
	\caption{(Color online) Snapshots of both the layers of the multiplex network where (a) 20\% and (b) 80\% inter-layer edges are heterogeneously delayed. The network parameters are $D_x = 5$, $E_y =0.2$. Other parameters are same as described in Fig.\ref{fig2}.}
	\label{fig7}
\end{figure}

 In addition, the extent of incoherent region of the chimera state depends upon the fraction of delayed inter-layer links $N_{\tau}$. The number of introduced heterogeneous delays perturb the same number of the mirror nodes in both the layers to produce the incoherent region. Fig.~\ref{fig7} (a) exhibits a chimera state with very small incoherent region arising due to small $N_{\tau}$ whereas Fig.~\ref{fig7} (b) exhibits a chimera state having a large incoherent region because of large $N_{\tau}$. This study demonstrates that besides regulating chimera state by varying network's structural parameters, the extent of incoherent region of the chimera state can also be regulated quantitatively by varying fraction of the multiplexing delays. \newline
\subsection{Designing the Chimera by multiplexing delays in Henon Map.-}
\begin{figure}[t]
	\centerline{\includegraphics[width=0.8\columnwidth]{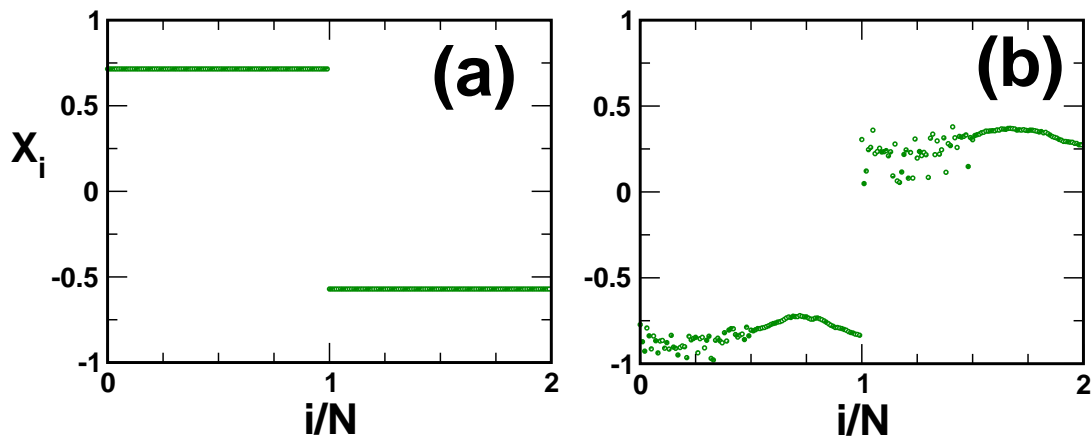}}
	\caption{(Color online) Snapshot profiles of the two layers of the considered multiplex network for (a) the undelayed interlayer links, (b) the delayed interlayer links with parameters $D_x=5, E_y=0.6$. The local dynamics is described by henon map (Eq.~\ref{hp_evoleq}) with parameters, $\alpha=1.4$ and $\beta=0.3$, $\varepsilon=0.9$, $\tau_{max}=20$, $r=0.32$, $N=100$ in each layer.
	}
	\label{fig8}
	\end{figure}

To verify if the regulating scheme is universally applicable, we also have investigated a multiplex network of non-locally coupled two-dimensional map, described as~\cite{hennon.map}
\begin{equation} \label{hp_evoleq}
	\begin{split}
		x_i^{t+1} &=f(x_i^t,y_i^t)+\frac{1}{(k_i)} \sum_{j=1}^{2N} C_{ij}[ f(x_j^{t-\tau_{ij}})-f(x_i^t,y_i^t) ] \\
		y_i^{t+1} &= \beta x_i^t
	\end{split}
	\end{equation}
	
where the local dynamics is governed by the Henon map $f(x_i^t,y_i^t)= 1 - \alpha (x_i^t)^2 + y^t$. As displayed in Fig.\ref{fig8}, two-dimensional Henon map also shows an engineered chimera state with delayed inter-layer links with proper choice of $D_x$ and $E_y$ parameter in the high coupling regime. Therefore, we deduce that the regulation scheme of engineering {chimeric pattern(s)} can be applied in variety of systems with different underlying dynamics provided the system lies in the {coherent regime}.
\section{Conclusion}
In the current study, we demonstrate a technique to produce {engineered} chimera states in a multiplex network by using any random initial condition in the presence of heterogeneous multiplexing (inter-layer) delays. We induce chimera states in the initially coherent multiplexed layers by introducing incoherence with the aid of multiplexing delays. It is also displayed that the emergent chimera {can be regulated} to one's choice both (a) qualitatively by tweaking the degree (level) of inducted incoherence by making proper choices for multiplex network's structural parameters such as interlayer and intralayer coupling strengths, and (b) quantitatively by tweaking the amount of inducted incoherence by varying fraction of the delayed interlayer links. 
The above-described {control} over the behavior of the emergent chimera can be understood in detail by the phase diagram in the interlayer and intralayer coupling parameters' space. The proposed scheme is {robust against underlying time-discrete local dynamics and might be applicable as well to continuous time} dynamical systems in producing engineered chimera states originating from regular initial conditions. {Also there may be cases when delays in the systems are inevitable and chimeras may not always be desirable. Such cases present a new challenge how naturally existing chimera in a delayed system can be destroyed and a modified application of the reported technique can be sought towards the cause.} 
This article sheds light on {manufacturing engineered} chimera in a multiplex network, whose relevance can be found in the case of neural disorders~\cite{chim.neuro.rev}.
The brain network possesses a highly complex structure of interconnections between neuronal cells~\cite{Neu_network}. The chemical and electrical synapses between them are responsible for most of our brain functions. Deterioration of these complex interaction pathways leads to the neural disorders which hinder brain from functioning normally. Interpreting this deterioration in individual neurons is a daunting task. Complex network approach presents an alternative holistic way to watch all the activities at once and find the anomalies in the emergent pattern within the framework of network science~\cite{Neu_dis_network}. Chimera is a promising candidate to detect these anomalous patterns in the brain. Furthermore, considering multiplex framework enables us to look even deeper into actual working of the constituents of the system which can not be achieved with a single layer network. 
Our technique of producing chimera with the aid of delays in the multilevel framework can provide a new direction in understanding the underlying dynamics behind the emergence of neuronal disorder in the brain as the delays are inherently present in the neuronal interactions inter-connecting different functional regions of the brain. \newline
\section*{Acknowledgments}.- SJ acknowledges DST project grant (EMR/2016/001921) and CSIR {project grant} (25 (0293)/18/ EMR - II) for financial support. SG {and ADK} acknowledges DST for the INSPIRE fellowship (IF150149) {and CSIR for RA position, respectively}. AZ acknowledges support by the Deutsche Forschungsgemeinschaft (DFG, German Research Foundation)—Projektnummer—163436311—SFB 910. We gratefully acknowledge the support from DST-DAAD PPP project {(INT/FRG/DAAD/P-06/2018)} which allowed us to have extensive collaboration and mutual visits. We thank Baruch Barzel for useful suggestions during his visit to IIT Indore.

\end{document}